# Comparison of block-based and hybrid-based programming environments in transferring programming skills to text-based environment

Hussein Alrubaye*, Stephanie Ludi†, Mohamed Wiem Mkaouer*
*Rochester Institute of Technology
†University of North Texas
{hat6622,mwmvse}@rit.edu, stephanie.ludi@unt.edu

*Abstract*—Teachers face several challenges when presenting the fundamental concepts of programming in the classroom. Several tools are introduced to give a visual dimension to support the learning process. These tools rely on code blocks, easily manipulated in a plug and play fashion, to build a program. These block-based tools intend to familiarize students with programming logic, before diving into text-based programming languages such as Java, Python, etc. However; when transitioning from block-based to text-based programming, students often encounter a gap in their learning. The student may not be able to apply block-based foundations in a text-based environment. To bridge the gap between both environments, we developed a hybrid-based learning approach. We found that on average a hybrid-based approach increases the students understanding of programming foundations, memorization, and ease of transition by more than 30% when compared to a block-based to text-based learning approach. Finally, we provide the community with an open source, hybrid-based learning tool that can be used by students when learning programming concepts or for future studies.

## I. INTRODUCTION

Teachers use different coding environments when teaching programming in the classroom. These coding environments are either block-based or text-based. Block-based approaches use blocks to write the program as introduced in Figure 1(A). Text-based approaches use text code only to write a program as shown in Figure 1(B). Tools such as PencilCode, Scratch, and App Inventor use a block-based approach. This environment is welcomed by millions of new students. App Inventor is used by 400,000 unique monthly active users who come from 195 countries and have created almost 22 million apps[1]. Scratch that is one of the most modern block-based development environment powered by MIT has more than 39 million users [2]. Furthermore, Block-based tutorials on code.org have been reaching over 780 million students [3]. This environment produces a new way to write code that includes colors and shapes. This can reduce the learning curve that students have when starting to learn to program in a text-based environment.

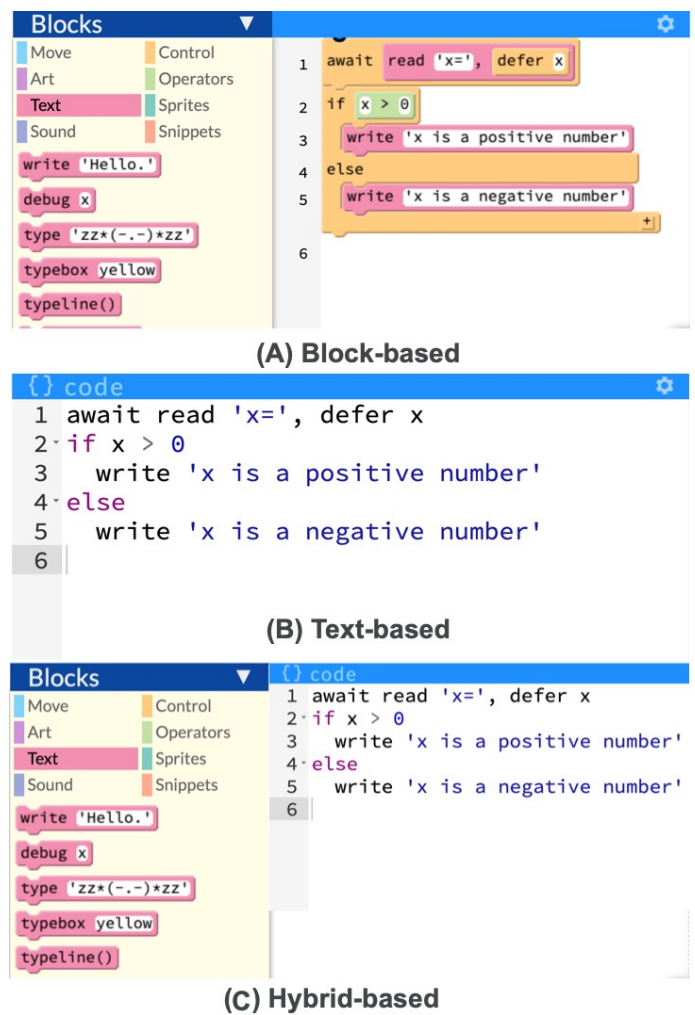

Figure 1: Types of programming environments.

It may seem more practical to leverage an existing reservoir of knowledge by extending the block-based approach towards a text-based one rather than starting to learn a whole new programming language [1]. However, there are some challenges teachers are facing to bring a text-based environment to the classroom. Teachers use block-based settings to teach programming because they are simple and easy to understand. However; with a block-based approach a student learns very little and is only able to write simple programs. Students will eventually need to move from a block-based approach to text-based approach in order to write complete and more complex programs. Students need to learn in a text-based environment in order to understand the difference between coding styles and coding syntax [1]. They need to transition from commands with colors and shapes to text-based environments with only commands. This transition includes large gaps in student learning and students are unable to transfer their skills in this new text-based environment [2].

We want to bridge the gap between block-based and text-based environments by implementing a hybrid-based environment, as proposed in Figure 1(C), which is a combination of block-based and text-based environments. This helps the learner in using block-based features while also being familiar with a text-based approach. It allows the learner to see and modify text-code while at the same time having the benefits of dragging and dropping blocks of code.

This study answers the following research questions:

- **RQ1. (Learning Improvement)** Does the hybrid-based environment better improve students learning curve, when they migrate to text-based environments, in comparison with block-based environment?

- **RQ2. (Command Memorization)** Does hybrid-based environment increase the student's memorization of programming commands, in comparison with block-based environment?

- **RQ3. (Ease Of Transition)** Does hybrid-based environment increase the ease of transition to text-based programming, in comparison with block-based environment?

The paper is structured as follows: Section II enumerates the studies relevant to our problem. Section III explains how we build our hybrid-based PencilCode tool. Section IV shows our experimental methodology in collecting the necessary data for the experiments that are discussed in Section V, followed by threads to validity in Section VI and the conclusion and future directions in Section VII.

## II. Related Work

The tools that we use in our learning have an impact on how and what we think. According to Dijkstra *"The tools we use have a profound (and devious!) influence on our thinking habits, and, therefore, on our thinking abilities."* [3]. Therefore, several studies focused on understanding the impacts of the development environment on learning curve [4], [5], [6], [7], [8], [9], [10] these studies show that the development environment could affect the learning curve. A student could perform differently in a different development environment. Other studies tried to bring visual programming to a high school classroom to help the student in the learning process [11], [12]. Also, other researchers study the impact of visual programming on mobile development by bringing the block-based mobile app in the classroom [13]. Various visual learning tools have been used to measure the impact of the development environment on the student learning curve.

Sherin [14] studied the learning of physics fundamental using either programming language or algebraic notation. They found that students who learned in different environments have different affordances in learning physics fundamental. Boroditsky [15] investigated the relationships between different environment representations and the learning curve. They found that different representations have different impact on student learning curve.

Weintrop [2] compared the impact of a block-based, hybrid- based and text-based on transfer programming skills using PencilCode. The study divided the classroom into three sections. The first section used block-based, the second section used text-based, and the third used hybrid-based (block /text). All students learned the same curriculum (ex, variables, loops, and conditions) for 5 weeks. In week 6, all students moved to text-based with java. They found that (92%) of students said that block-based is easier to learn programming than text-based. Because it is easy to drop and fewer memories commands also students had a better attitude to understand the loops, variables than text-based students. However, the students found that the drawback for block-based is challenging to build a large complex program. The hybrid-group did not have their tool. They are switching between code and block which make it difficult for students to track the representation of blocks in the code when the program becomes large. So they did not get fair learning comparison. They have to have their tool.

Robinson [1] did a study on Scratch. One of an exciting feature in Scratch is simple to understand and use with avoiding the learner the syntax errors. While Scratch focuses on learning the programming logic, not programming languages. The student does not learn how to build a program. Instead, the student learns how to think logically [1]. When student transfer from Scratch to the text-based environment, he/she does not have any programming background.

In this paper, we perform a comparative study between hybrid-based approach and block-based approach in the context of transferring programming skills to text-based.

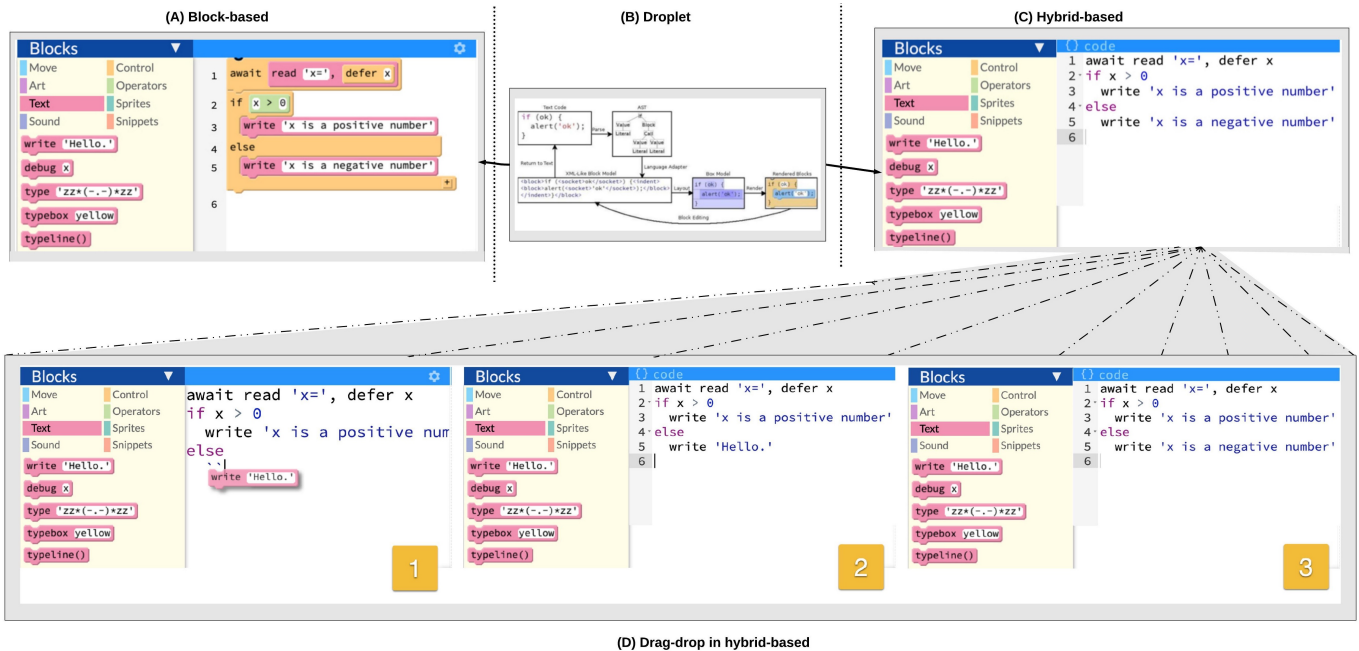

Figure 2: Hybrid-based PencilCode Overview.

## III. Methodology

In this section, we explain how we developed a hybrid-based tool using PencilCode. We start by analyzing the architecture design of the block-based PencilCode. We then discuss the modifications that we made to the block-based PencilCode in order to transform it into a more hybrid-based PencilCode. Figure 2, represents our hybrid-based PencilCode system architecture.

We have chosen to work with PencilCode because it helps to build confidence in beginning programmers so that they can create more complex programs without using a block-based approach [16]. Also, it allows beginners to achieve satisfactory results quickly, while also minimizing their frustration level when learning, by avoiding syntactic errors that can be easily introduced when typing down instructions. Furthermore, it introduces beginners directly to a programming foundation that is used by professionals and allows them to toggle between text-based and block-based environments [16]. PencilCode enables students to write a real CoffeeScript program using blocks only. Figure 2(A) shows an example of writing code in PencilCode. PencilCode is currently open source, and available in GitHub[4].

In PencilCode, a user can switch between text-based and block-based by clicking on "*show code*" or "*show block*". When the user clicks on "*show code*", the blocks view transitions from Blocky[5] to a Droplet [17] model and then the code is displayed. When the user clicks on "*show block*" the code view transitions from a Droplet to a Blocky model and then the blocks are displayed. This illustrates that PencilCode uses two different models, a Blocky model and a Droplet model. All reserved commands (For, IF, variables, etc.) are available to users in the toolbox as a block so that they can drag and drop (for convenience, we will just say drop) when building an application in the Blocky view.

Figure 2(B) shows how a Droplet is used to convert blocks to code and code to blocks. When a user drops blocks from the toolbox to the text-based view, the Droplet block-model displays its corresponding program as a number of connected blocks. To build a hybrid-based tool, we need to convert the block-model into its corresponding textual representation.

We build the hybrid-based PencilCode to reduce the learning gap between the block-based and text-based approaches as outlined in Figure 2(C). The user is able to use the blocks while also seeing the code. This increases the liaison between the block representation and its corresponding code representation, and also allows the user to learn the syntax of programming. In hybrid-based PencilCode a user writes an application by following these steps:

- First, the user drags the block from the blocks toolbox, as shown in Figure 2(D- 1).
- Second, the user drops the block into text-based, as shown in Figure 2(D- 2).
- Third, the user can then either use the text-based or update the text, as shown in Figure 2(D- 3).

In the next section, we discuss the details of updating in PencilCode to build hybrid-based PencilCode.

---

## A. Droplet customization

In order to build a hybrid-based tool, we first need to modify the design of the Droplet model in PencilCode. The Droplet is used as part of the PencilCode tool when switching from a block-based to a text-based model. The Droplet's data model is a text stream marked up with XML-like tokens such as $< block >, < /block >, < socket >, < /socket >$, where every block that is dropped to the Droplet editor is marked as a token with a start tag and an end tag.

To convert the text code to blocks, instructions are parsed and converted to an Abstract Syntax Tree (AST) (see Figure 2(B)). The parent node along with its child nodes, all the way to leaf nodes are then identified. This tree is then converted to an XML-like block-model using a syntax-aware language adapter that is responsible for mapping the block-model to the syntactic code representation, as follows: The Droplet needs to render the XML box-model to blocks. The Droplet parses the block-model line by line and, for each line, it transforms the text between two markups into a block.

For better visibility, we also defined a set of rules to insert spaces between lines in each box-model. These spaces make the block clearer when it is drawn. The Droplet draws the path that surrounds all of its rectangles while avoiding the unintended area.

The Droplet typically displays an animation when converting any text to a block. We disable this animation in the hybrid-based environment as it would lead to a large number of frequent animations and thus confuse developers as they are writing their instructions and being constantly distracted by an animation.

The hybrid-based PencilCode converts every dropped block to code instantly. This allows students to see how every block is represented in the code, as they are developing their program. We send every block to the Droplet model and then convert it to code instead of sending the entire program and then converting it to box-model (text-based). Students can then edit and write code directly in the Droplet editor.

## B. Hybrid-based PencilCode implementation

hybrid-based PencilCode, as shown in Figure 1(C), uses the same internal design of PencilCode with some adjustments. We intentionally avoided architectural changes, and minimized code changes to keep our extension easy to implement and maintain. Also, this would allow more compatibility with any upcoming version of PencilCode. Table I contains a high-level overview of all our updates and their corresponding PencilCode files. In each file, we discuss the rationale and details of our updates:

In **view.js** we made two changes: *First*, Instead of allowing students to click on *"show code"* and *"show block"* to switch between block-code, we automated this process implicitly, without letting the student notice it.

To do that, we added a new function that is being called whenever a student drops a block from the toolbox to the text-based area or whenever the student updates the text-based code.

When a student drops a block, the function passes the block to the Droplet in order to generate the code, that is now instantly visible to the student. This enables the code view of each block. Furthermore, when a student changes a line of code, this is a captured event that updates the blocks with respect to the updated code.

*Second*, In hybrid-based PencilCode, both block-based (toolbox only) and text-based are viewed in front of the user instead of one at the time. This view is achieved by a minor update in Style Sheets, which manage program blocks or code views. We retire PencilCode's method that handles the correspondence between blocks and instructions. So when a student drops a block, our Droplet handles that event by generating the necessary text-based information, and when a line of code is modified, our Droplet also updates the blocks.

In **Droplet.js**, we made four changes, as enumerated in Table I. These changes allow the Droplet to support converting block to code instead of block to block only. When a student drops a block in the text-based area, these changes convert the block to its corresponding source code.

Other changes that were made in **editor.html**, and **Droplet.css** are related to user interface design. In order to combine both block-based and text-based views into one uniform hybrid-based environment. We modify the layout code of PencilCode from switching between two environments which are the block-based environments and the text-based environments into one environment which is the hybrid-based environment. The user interface of the hybrid-based environment IDE splits into two views as elicited in Figure 1(C). The left side is toolbox where a student uses to drag a programming command, and the right side is the development area(text-based) where student drop blocks. We named hybrid-based environment as *hybrid-based PencilCode* that released as open source project[6].

Table I: Changes in PencilCode code files to build hybrid-based PencilCode.

| File Name | File Path | Line numbers | Operation |
|-----------|-----------|--------------|-----------|
| editor.html | content/ | 13-15 | update |
| Droplet.css | content/lib/ | 49-51 | add |
| Droplet.js | content/lib/ | 65308- 65310 | update |
| | | 65479- 65480 | update |
| | | 65826- 65829 | update |
| | | 65308- 65310 | remove |
| view.js | content/src/ | 2841-2842 | remove |
| | | 1933-1943 | add |

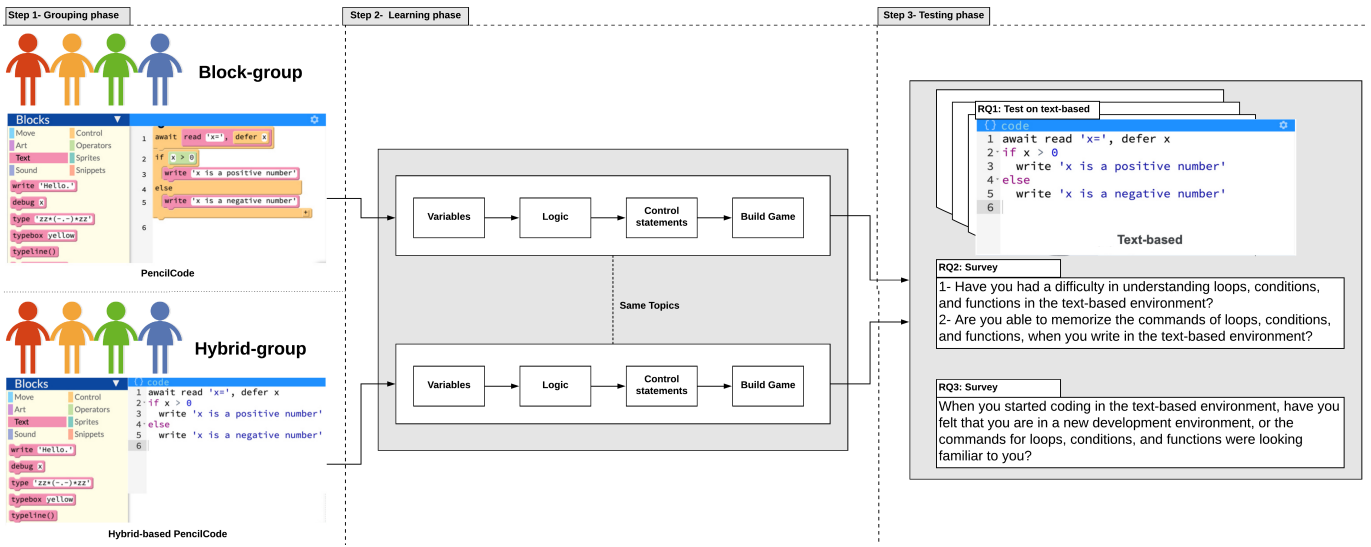

Figure 3: Experimental Design Overview.

## IV. Experimental Design

We design our experimental study to measure the difference in the impact of block-based or hybrid based environment on student's learning curve, when they transition to the text-based environment. To do so, we perform a qualitative analysis of two separate groups of students, where each group is assigned to only one environment (block-based or hybrid-based). Then we evaluate the effect of these environments on transferring the basic programming skills to a more complex environment, i.e., text-based programming. We want to investigate whether our hybrid-based environment outperforms the classic block-based environment in terms of optimizing the learning time and reducing the error-proneness.

We design our experiment in three phases, as outlined in Figure 3. First, in *Grouping phase* we divide students into two groups: the *block-group* and the *hybrid-group*. Second, in *Learning phase* we teach each group basic programming concepts using its associated environment, namely block-based and hybrid-based. Third, in *Testing phase* we perform a test, in the text-based environment, to challenge the students understanding of programming concepts, and finally we survey them to gauge their impression of the ease of programming in general.

### A. Grouping phase

As shown in Figure 3 (step 1), eighteen undergraduate students from the civil and environmental engineering departments were randomly sampled for this study. We verified that they have no prior experience in programming and every team has( 4 males, and 5 females) . They were hired for two sessions, of 2 hours each. We randomly divided the students into two equal groups: (1) *Block-group*, this group learns programming using block-based PencilCode; and (2) *Hybrid-group:* this group learns programming using hybrid-based PencilCode. For the learning phase, we scheduled separate sessions for each group, and we did not disclose their existence to each other, in order to avoid any communication between teams, in terms of sharing materials or questions, and this that may affect the accuracy of our experiments and results.

### B. Learning phase

As shown in Figure 3 (step 2), the Block-group learns programming using the block-based PencilCode environment while the hybrid-group learns programming using the hybrid-based PencilCode environment. We scheduled to teach materials in basic foundations of programming, including variables, conditions, and loops. Then we built easy-to-program games [7]. We teach both classes using the same material, so we can ensure the fairness between both environments. We use the projection of materials and we allow students to apply programming topics in a by-Example fashion. This allows better visualization of concepts as we demonstrate the execution of every program that we teach during the sessions.

### C. Testing phase

As shown in Figure 3 (step 3), After teaching every group how to write programs in their corresponding environment, in this phase, both teams transition to the text-based environment, where we have prepared a common test for both groups along with a survey for all participants. Our experiments are driven by the previously stated research questions.

To answer **RQ1. (Learning Improvement)**, we performed three types of evaluations: (1) *Code Modification*,

---

[7]All materials are in attachment supplementary materials

we test the student's ability in correcting a faulty text-based program. We give students a buggy code and its related correct input/output, then we ask them to locate the root cause and fix it. For example, the code below checks if a particular number is positive or negative.

---

**Test sample 1: Is 'x' value positive or negative**

```
x = 7
if x >0
    write 'x is a positive number.'
else
    write 'x is a negative number.'
```

---

We ask the students to update the code and handle the case of the unsigned number "0". Students are required to take into account the case "x=0". (2) *Syntax Error Free*, we test the student's ability in finding syntax error(s). For example, In the question below, we added a syntax error in the condition statement by making the indent of the condition and condition block of code line starting at same point. This represents an error in CoffeeScript, because it is a space sensitive language, therefore, the body of the condition statement has to be indented.

---

**Test sample 2: What is the output**

```
sum=0
for x in [0..10]
    if x>8
    sum=sum+x    //<---------- Syntax Error
write 'sum= '+ sum
```

---

Figure II shows multiple-choice answers for the possible output in test sample 2. students need to select one answer that they think it is correct. In this case, the correct answer is the option (C). Furthermore, we consider any syntax error that has been introduced by the student in any of their code updates as a valid value to calculate syntax error free matrix.

Table II: Possible outputs for test sample 2.

| A | B | C | D |
|---|---|---|---|
| sum= 9 | sum= 19 | not run | sum= 9 <br> sum= 19 |

(3) *Ease Of Learning*, we test the student's ability in deciphering and understanding the code logic. For example, as written in the listing below, we ask students: for the following code what do you think the tortoise will draw?

---

**Test sample 3: Tortoise movement**

```
speed 2
pen red
for [1..10]
    fd 100
    rt 45
```

---

Figure 4 shows multiple-choice answers for the possible shapes that tortoise may draw. Students need to select one answer that they think it is correct. In this case, the correct answer is the option (D), and not (B), since tortoise actually makes ten moves.

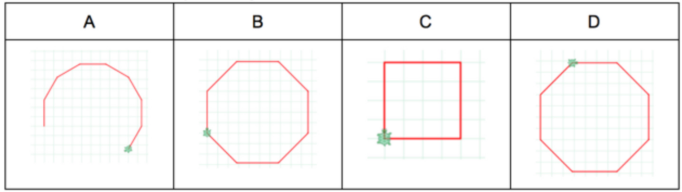

Figure 4: Possible shapes that a tortoise may draw.

We designed an overall of 25 different questions in code modification, syntax error free, and ease of learning. To guarantee the pedagogical aspect of these questions, we have mainly selected them from Weintrop's study [2]. All the questions that used in this study are available online[8]. The grading scale varies between 0 (bad) and 100 (good).

To answer **RQ2. (Command Memorization)**, we survey students using the following questions:

1) *Have you had a difficulty in understanding loops, conditions, and functions in the text-based environment?*
2) *Are you able to memorize the commands of loops, conditions, and functions, when you write in the text-based environment?*

Both questions are answered using a Likert scale [18], [19], varying between 0 (bad) and 5 (good).

To answer **RQ3. (Ease Of Transition)**, we survey students using the following question:

1) *When you started coding in the text-based environment, have you felt that you are in a new development environment, or the commands for loops, conditions, and functions were looking familiar to you?*

Both questions are answered using a choice of either *"Yes, it looks new"(0 star)* or *"No, it looks familiar"(5 stars)*. We opted for a binary answer to capture student's decisiveness of whether they are comfortable or not, with the text-based environment.

In the next section, we discuss the qualitative analysis for the tests and surveys results.

---

[8]All questions are in attached supplementary material

## V. Results

### A. Results for RQ1. (Learning Improvement)

To answer our first research question, we report, in Figure 5, the results of grading both groups in their ability to modify code, debug it and comprehend it. Figure 5 contains each group average grade for *Code Modification*, *Syntax Error Free*, and *Ease Of Learning*.

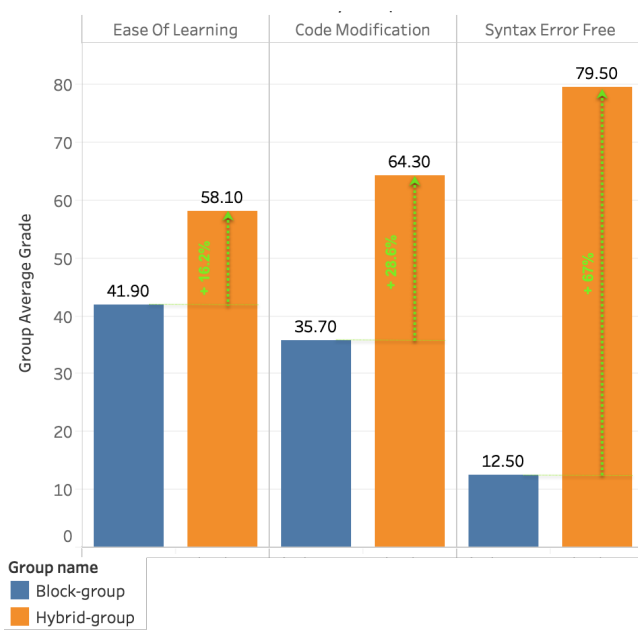

Figure 5: Performance of hybrid and block learning improvements (higher is better).

For **Code Modification**, the hybrid-group has an average of 64.30%, while the block-group has scored an average of 35.70%. Thus, students belonging to hybrid-group have experienced a higher ability in correctly modifying the code in the text-based environment, in comparison with the students of the block-group. Furthermore, a Mann-Whitney $U$ test, between the difference of grades between the two groups has shown significance ($p$-value $\leq 0.05$). We note that, although block-based environments have various advantages in facilitating programming concepts, they do limit the learner's early exposure to the actual source code, like in high-level programming languages [17], which hinders their ability to discover syntactic errors. This explains the difficulty experienced by the block-group in capturing logical errors. On the other hand, the hybrid environment facilitates the early interaction between beginners and the source code, in a way that allows updating their code from both, block and source code views.

For **Syntax Error Free**, we observe that students who learned with a block-based environment have a higher probability of producing syntax errors in the text-based environment, when compared with the hybrid-group students. As depicted in Figure 5, on average, 79.5% of the hybrid-group's students write instructions that are free of syntactic errors. However, only 12.5% of the block-group's students write code that is free of syntactic errors. Also, the difference in the number of errors of each student, clustered by their group, is significant ($p$-value $\leq 0.05$). These results highlight the importance of early raising the awareness of beginners to the syntactic nature of programming in general. Being inline with this concept, our proposed hybrid-based environment views the basic syntax properties as part of the translation from blocks to source code. For instance, students discover spacing in CoffeeScript, while they write their hybrid-based program.

As for **Ease Of Learning**, we note from Figure 5, that hybrid-group students score an average grade of 58.10%, while block-group students score an average grade of 41.90%. Also, we report the statistical significance of the difference between the two sets of grades ($p$-value $\leq 0.05$). Thus, we report that our hybrid-based environment improves the student's learning by 16.2% in comparison with the block-based environment. More concretely, hybrid-group students can drop blocks of code without the need to memorize the commands. At same time, they repeatedly observe how the block is converted to its corresponding source code when they drop blocks from the toolbox to text-based view. While students in the block-group could only see blocks in their views and development area.

> To summarize our findings, we observe that the hybrid-based environment has improved the students learning curve when migrating to the text based environment. Students using the hybrid-based environment are also able to effectively debug the code from seeded errors, outperforming students using the block-based environment by 28.6% on average. Furthermore, the percentage of students with error-prone source code is 67% less in the hybrid-group, in comparison with the block-based group. Finally, the hybrid-group outperforms the block-group in identifying the programming concepts by 16.2%

### B. Results for RQ2. (Command Memorization)

When students, from both groups, are writing coding in the text-based environment, 54.5% (2.725/5) of the hybrid-group's students were able to memorize programming commands while only 45.5% (2.275/5) of the block-group's students were successful in memorizing programming commands. As a result, the hybrid-group is 9% better than block-group in command memorization. Students in the hybrid-group were exposed to modifying commands, as part of their environment. Besides, they see every programming command as a block in the toolbox. As a result, seeing commands as blocks and being able to drop and change them, leads to a better grasp of the commands.

> As a summary, hybrid-based environment is 9% better than the block-based environment in the memorization of programming commands.

### C. Results for RQ3. (Ease Of Transition)

We found that the hybrid-group has a smoother transition to the text-based environment than the block-group, as shown in Figure 6. When students of both groups are writing code in the text-based environment, 80% (4/5) of the hybrid-group's students expressed a noticeable ease of transition to text-based, by answering with *"No, it looks familiar"*. While 50% (2.5/5) of the block-group's students expressed an ease of transition to text-based, as they have chosen the second answer. According to the survey results, hybrid-group is 30% confident than block-group about programming in a text-based environment. Practically, students of the hybrid-group are in touch with the code while they learn how to program, in contrast with block-group students, who found the text-based environment to be new to them.

> As a summary, learning in hybrid-based environment, increases confidence by 30% better than the block-based environment, in the ease of the transition to in programming in text-based environment.

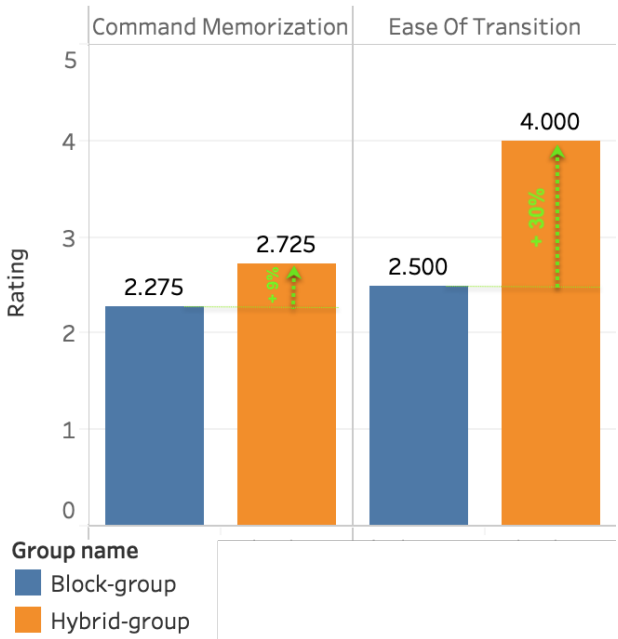

Figure 6: Survey results of hybrid and block learning improvements (higher is better).

### VI. THREATS TO VALIDITY

Our study inherits threats that are related to studies of students and programming languages. First of all, our code examples were specific and may not necessarily be representative of all programming concepts. To mitigate this, we test students on the same concepts they have been exposed to, during the learning phase, besides relying on questions, used in previous studies [2], and both quantitative and qualitative analyses to enhance the accuracy of our observations.

Another threat is related to the random division of students into groups, in which we cannot guarantee a uniform distribution of learning skills across groups. However, it is eventually challenging to estimate the programming learning skills of any student with no programming background. To reduce the bias in learning skills, we verified that the students have no learning background by checking their degrees and the courses they have taken in their academic career. Also, to reduce the sample bias, we have chosen students from different levels (freshman, sophomore, etc.), and belonging to various degrees, as long as they have no exposure to programming.

Another factor that may influence the observed results, is the lack of interest of some students during the testing phase, which may increase their proneness to errors. prior to them any their learning skills. To mitigate this risk, we only hired students who have expressed interest to learning programming and we also paid them $60 upon the completion of their task along with extra $40 for those who achieved no errors to motivate them. Also, we performed this study on a limited timeline, results would be more accurate if the experiment is performed throughout a longer period to allow students with slower a learning curve to better capture the concepts.

### VII. CONCLUSION

In this paper, we present a novel approach in bridging the gap between block-based and test-based programming environments, through merging them in one hybrid environment. The qualitative analysis of our proposed approach as shown promising results in terms of improving students learning curve by an average of 30.16% (learning programming foundations improved by 16.2%, learning code modification improved by 28.6%, command memorization by 9%, error free code by 67%, and ease of transition by 30%), when compared to the block-based environment. Furthermore, We implement a hybrid-based version of PencilCode. Our tool is open source[9] for learning, replication, and extension purposes.

#### ACKNOWLEDGMENT

We sincerely thank David Bau who is the author of the PencilCode that we have used in this study, for supporting and for providing PencilCode as open source.

# References

[1] William Robinson. From scratch to patch: Easing the blocks-text transition. In *Proceedings of the 11th Workshop in Primary and Secondary Computing Education*, pages 96–99. ACM, 2016.

[2] David Weintrop. Blocks, text, and the space between: The role of representations in novice programming environments. In *2015 IEEE Symposium on Visual Languages and Human-Centric Computing (VL/HCC)*, pages 301–302. IEEE, 2015.

[3] Edsger W Dijkstra. How do we tell truths that might hurt? In *Selected Writings on Computing: A Personal Perspective*, pages 129–131. Springer, 1982.

[4] David Weintrop. *Modality matters: Understanding the effects of programming language representation in high school computer science classrooms.* PhD thesis, Northwestern University, 2016.

[5] Wallace Feurzeig et al. Programming-languages as a conceptual framework for teaching mathematics. 1971.

[6] Michael S Horn and Robert JK Jacob. Designing tangible programming languages for classroom use. In *Proceedings of the 1st international conference on Tangible and embedded interaction*, pages 159–162. ACM, 2007.

[7] Hsin-Kai Wu, Joseph S Krajcik, and Elliot Soloway. Promoting understanding of chemical representations: Students' use of a visualization tool in the classroom. *Journal of Research in Science Teaching: The Official Journal of the National Association for Research in Science Teaching*, 38(7):821–842, 2001.

[8] Marcia C Linn. The cognitive consequences of programming instruction in classrooms. *Educational Researcher*, 14(5):14–29, 1985.

[9] Yota Inayama and Hiroshi Hosobe. Toward an efficient user interface for block-based visual programming. In *2018 IEEE Symposium on Visual Languages and Human-Centric Computing (VL/HCC)*, pages 293–294. IEEE, 2018.

[10] Arjun Rao, Ayush Bihani, and Mydhili Nair. Milo: A visual programming environment for data science education. In *2018 IEEE Symposium on Visual Languages and Human-Centric Computing (VL/HCC)*, pages 211–215. IEEE, 2018.

[11] Emmy-Charlotte Förster, Klaus-Tycho Förster, and Thomas Löwe. Teaching programming skills in primary school mathematics classes: An evaluation using game programming. In *2018 IEEE Global Engineering Education Conference (EDUCON)*, pages 1504–1513. IEEE, 2018.

[12] D Midian Kurland, Roy D Pea, Catherine Clement, and Ronald Mawby. A study of the development of programming ability and thinking skills in high school students. *Journal of Educational Computing Research*, 2(4):429–458, 1986.

[13] Don Kerrigan. Creating a classroom programming lab using android and blockly. In *Proceedings of the 49th ACM Technical Symposium on Computer Science Education*, pages 275–275. ACM, 2018.

[14] Bruce L Sherin. A comparison of programming languages and algebraic notation as expressive languages for physics. *International Journal of Computers for Mathematical Learning*, 6(1):1–61, 2001.

[15] Lera Boroditsky. Does language shape thought?: Mandarin and english speakers' conceptions of time. *Cognitive psychology*, 43(1):1–22, 2001.

[16] David Bau, D Anthony Bau, Mathew Dawson, and C Pickens. Pencil code: block code for a text world. In *Proceedings of the 14th International Conference on Interaction Design and Children*, pages 445–448. ACM, 2015.

[17] David Bau. Droplet, a blocks-based editor for text code. *Journal of Computing Sciences in Colleges*, 30(6):138–144, 2015.

[18] Rensis Likert. A technique for the measurement of attitudes. *Archives of psychology*, 1932.

[19] John Brooke et al. Sus-a quick and dirty usability scale. *Usability evaluation in industry*, 189(194):4–7, 1996.